\documentclass[12pt]{article}
\usepackage{graphics}
\usepackage{amssymb}

\begin{document}

\parskip=2pt
\parindent=7mm
\renewcommand{\baselinestretch}{1.}
\renewcommand{\theequation}{\arabic{section}.\arabic{equation}}

\newcommand{\be}{\begin{equation}}
\newcommand{\ee}{\end{equation}}
\newcommand{\ba}{\begin{eqnarray}}
\newcommand{\ea}{\end{eqnarray}}
\newcommand{\bd}{\begin{description}}
\newcommand{\ed}{\end{description}}
\newcommand{\pd}{\partial}
\newcommand{\CC}{{\mathcal{C}}}
\newcommand{\A}{{\mathbb{A}}}
\newcommand{\B}{{\mathbb{B}}}
\newcommand{\C}{{\mathbb{C}}}
\newcommand{\HHH}{{\mathbb{H}}}
\newcommand{\R}{{\mathbb{R}}}
\newcommand{\V}{{\mathbb{V}}}

\title{Geometric phase related to point-interaction transport
on a magnetic Lobachevsky plane}
\date{}
\author{S.A.~Albeverio,$\!^{a}$ P.~Exner,$\!^{b,c}$
V.A. Geyler$^{d}\!$}
\maketitle

\begin{quote}
{\small {\em a) Institute for Applied Mathematics, Universit\&quot;at
Bonn, \\ \phantom{a) }Wegelerstr.~10, 53115 Bonn, Germany} \\ {\em
b) Nuclear Physics Institute, Academy of Sciences, 25068 \v Re\v z
\\ \phantom{b) }near Prague, Czechia
 \\ c) Doppler Institute, Czech Technical
University, B\v rehov{\'a} 7,\\ \phantom{a) }11519 Prague,
Czechia}
 \\ {\em d) Department of Mathematical Analysis, Mordovian State \\
 \phantom{a) } University,  430000 Saransk, Russia;} \\
 \phantom{a) }\texttt{albeverio@uni-bonn.de} \\
 \phantom{a) }\texttt{exner@ujf.cas.cz},
 \texttt{geyler@mrsu.ru} }
\end{quote}

\begin{abstract}
\noindent We consider a charged quantum particle living in the
Lobachevsky plane and interacting with a homogeneous magnetic
field perpendicular to the plane and a point interaction which is
transported adiabatically along a closed loop $\CC$ in the plane.
We show that the bound-state eigenfunction acquires at that the
Berry phase equal to $2\pi$ times the number of the flux quanta
through the area encircled by $\CC$.
\end{abstract}


\section{Introduction}

The phenomena arising from a geometric phase called Berry phase
\cite{Ber} have been put in evidence in many quantum mechanical
systems. Recently such a ``Berry phase effect'' has been observed
in some magnetic systems \cite{LSG,MHK}. Moreover, it was shown
that the geometric phase can emerge even in a time-independent
homogeneous magnetic field when a potential well trapping a
two-dimensional particle is transported along a closed loop. An
example in which the potential is of zero range is worked out in
\cite{EG}, where a formula was proved showing that in the absence
of an additional confining potential the acquired phase is
proportional to the number of flux quanta through the area
encircled by the loop.

This result raises the natural question whether it can be extended
to systems with a nontrivial configuration-space geometry. In this
letter we address this problem in the framework of a solvable
model in which the Euclidean plane is replaced by a Riemannian
manifold of a constant negative curvature; we shall show that the
analogue of the mentioned ``planar'' formula is valid here. Among
other things, our result illustrates an important difference
between two kinds of geometric phases, namely those of Berry and
of Aharonov--Anandan. Specifically, the Aharonov--Anandan
connection is closely related to the metric connection of the
parameter spaces ${\bf C}P^N$ \cite{Mos}, whereas the Berry
connection is completely independent of the Levi--Civita
connection in the parameter space.


\setcounter{equation}{0}
\section{The free Hamiltonian}

The configuration space of our model is the Lobachevsky plane,
i.e. a complete two-dimensional simply connected Riemannian
manifold of constant negative curvature $R, \, R<0$. We shall
employ the Poincar\'{e} realization in which the Lobachevsky plane
is identified with the upper complex halfplane
$$ \HHH^2_a = \{ z\in\C:\; \Im z>0 \} $$
endowed with the metric
$$ ds^2 = {a^2\over y^2}\, (dx^2 +dy^2)\,, $$
where $x=\Re z,\, y=\Im z$, and $a>0$ is the parameter related to
the curvature $R$ by $R=-2/a^2$. Then the geodesic distance on
$\HHH^2_a$ is given by the formula
$$ d_a(z,z') = a\: \mathrm{Arcosh\,} \left\lbrack 1+\,
{|z-z'|^2\over 2yy'} \right\rbrack $$
and the area element $d\mu_a$ has the form
\be d\mu_a = {a^2\over y^2}\, dx\wedge dy\,. \label{mu} \ee
A constant magnetic field on $\HHH_a^2$ is given by a 2-form $\B$
defined as
$$ \B = {Ba^2\over y^2}\: dx\wedge dy\,, $$
where $B$ is the field intensity. The form $\B$ is obviously exact
and any 1-form $\A$ such that $\B= d\A$ is called a vector
potential related to the field $\B$. For our purpose it is
convenient to choose $\A$ in the Landau gauge
$$ \A = {Ba^2\over y}\, dx\,. $$

The Schr\"odinger operator describing a particle of charge $e$ and
mass $m_*$ which lives on the Lobachevsky plane $\HHH^2_a$ and
interacting with a magnetic field is according to \cite{Com} given
by
\be H^0 = -\, {\hbar^2\over 2m_* a^2}\, \left\{ y^2 \left(
{\pd^2\over\pd x^2} + {\pd^2\over\pd y^2} \right) - 2iby
{\pd\over\pd x}\, - b^2 \right\} -\,{\hbar^2 \nu\over 8m_* a^2}
\,, \label{freeham} \ee
where we have introduced the dimensionless quantity
$$ b = {eBa^2\over \hbar c} $$
which has a simple meaning: if $\Phi_e= {2\pi\hbar c\over e}$ is
the magnetic flux quantum relative to the charge $e$, then $b$ is
the doubled number of flux quanta through the degenerate triangle
(the area of which is $\pi a^2$).

The presence of the last term on the r.h.s. of (\ref{freeham})
(this term is absent in \cite{Com}) can be justified in different
ways, e.g. as a ``van Vleck correction'' \cite{Gut}, in which case
$\nu=1$. On the other hand, in one embeds locally the Lobachevsky
plane into $\R^3$ and derive the Hamiltonian through a squeezing
limit of a saddle-shaped layer \cite{Tol} one obtain the last term
with $\nu=4$. We shall not discuss this difference, however,
because it is of no importance for the result we are going to
derive in this paper. For the sake of simplicity we put in the
following $e=c=\hbar=2m_*=1$.

The spectrum of $H^0$ consists of two parts \cite{Com}, the second
one being absent for weak fields, $2|b|\le 1$:
\bd
 \item{\em (i)}
 an absolutely continuous spectrum in the interval
 $[b^2/a^2,\infty)$,
 \item{\em (ii)}
 a point spectrum consisting of a finite number
 of infinitely degenerate eigenvalues $E^0_n,\:
 0\le n< |b|-{1\over 2}\,$. These Landau levels are given
 explicitly:
\be E^0_n = {1\over a^2} \left( b^2\! -\! \left(|b|\!-\!n\!-
{1\over 2}\right)^2 \right) = {1\over a^2} \left( |b|(2n\!+\!1) -
\left(n\!+ {1\over 2}\right)^2 \right)\,. \label{ev} \ee
\ed
We will need also an explicit expression for the Green's function
$G^0(z,z';\zeta)$ of $H^0$. Let us introduce the quantity
\be \sigma(z,z') = \cosh^2\left(d_a(z,z')\over 2a \right)\,.
\label{sig} \ee
It is easy to see that it is independent of $a$ being equal to
$$ \sigma(z,z') = {|x\!-\!x'|^2 + |y\!+\!y'|^2 \over 4yy'}  $$
Given $\zeta \in \C\setminus [ b^2/a^2, \infty)$ we put
$$ t(\zeta) = {1\over 2}\,+\, \sqrt{b^2\!-a^2\zeta}\,, $$
where the square root $\sqrt z$ is defined in the cut plane
$\C\setminus (-\infty,0)$ by the requirement $\Re\sqrt{z}\ge 0$.
With this notation the integral kernel of $(H^0\!-\zeta)^{-1}$ is
of the form \cite{Els}
$$ G^0(z,z';\zeta) = {1\over 4\pi} \left(- {z\!-\!\bar z' \over
\bar z \!-\!z'} \right)^b {\Gamma(t\!+\!b) \Gamma(t\!-\!b) \over
\Gamma(2t)}\; \sigma^{-t}\, F(t\!+\!b, t\!-\!b; 2t;
\sigma^{-1})\,, $$
where $F(a,b;c;x)$ is the hypergeometric function.


\setcounter{equation}{0}
\section{Krein's formula}

Now we shall consider a point perturbation -- introduced in the
usual way \cite{BF, AGHH} -- of the operator $H^0$ supported by a
point $w\in \HHH^2_a,\: w=u\!+\!iv$. By Krein's formula
\cite[App.~A]{AGHH} the Green's function of the perturbed operator
has the form
\be G(z,z';\zeta) = G^0(z,z';\zeta) - {G^0(z,w;\zeta)
G^0(w,z';\zeta) \over Q(\zeta)-\alpha}\,, \label{krein} \ee
where the parameter $\alpha$ is related to the scattering length
$\lambda$ of the point ``potential'' by the formula
$$ \alpha = {m_*\over \pi\hbar^2}\, \ln\lambda $$
(or $2\pi\alpha=\ln\lambda$ in the rational units). The Krein's
function $Q(\zeta)$, defined as the regularized trace of the
free-resolvent kernel, was evaluated in \cite{BG} to be
$$ Q(\zeta) = {1\over 4\pi}\, \left\lbrack \psi(t\!+\!b) +
\psi(t\!-\!b) + 2\gamma - 2\ln 2a\, \right\rbrack\,, $$
where $\psi(z)$ is Euler's digamma function and $\gamma=\psi(-1)=
0.577...$ is the Euler number. The perturbed Hamiltonian with the
Green's function (\ref{krein}) will be denoted as $H_{w,\alpha}$.

Using the well-known behaviour of the digamma function \cite{AS,
BE} we find that the Krein's function has in our case the
following properties:
\bd
\item{\em (a)} $\,Q(\zeta)$ is a meromorphic function in the cut
plane $\C\setminus [b^2/a^2,\infty)$ with the poles at the points
$E^0_n$ of the discrete spectrum of $H^0$ given by (\ref{ev}).
\item{\em (b)} $\, \lim_{\R\ni\zeta\to -\infty} Q(\zeta) =
-\infty$.
\item{\em (c)} At the continuum threshold we have
$$ \lim_{\R\ni\zeta\to b^2/a^2} Q(\zeta) = \left\{
\begin{array}{ccl} +\infty & \quad\dots\quad & |b|\;\:
{\rm half-integer} \\ q_{a,b} & \quad\dots\quad &
\mathrm{otherwise}
\end{array} \right. $$
where $q_{a,b}= {1\over 4\pi}\, \left\lbrack \psi({1\over 2} +\!b)
+ \psi({1\over 2}-\!b) + 2\gamma - 2\ln 2a\, \right\rbrack $.
\item{\em (d)} $\,{\pd Q\over\pd\zeta} >0$ holds at each point of
$\R\setminus\sigma(H^0)$.
\ed
As usual we employ the symbol $[x]$ for the integer part of a
number $x$. We set $n_0= \lim_{\varepsilon\to 0} \left\lbrack |b|-
{1\over 2}- \varepsilon \right\rbrack$ and consider the following
family of intervals
$$ (-\infty,E^0_0),\: (E_0^0,E_1^0),\: \dots,
(E_{n_0-1}^0,E_{n_0}^0),\: (E_{n_0}^0,b^2/a^2)\,. $$
We also set $E^0_{-1}=-\infty$ so for $n_0=-1$ the family consists
of a single interval $(-\infty,b^2/a^2)$. The last interval having
$b^2/a^2$ as the right endpoint will be called special, while all
the other intervals are dubbed regular.

The listed properties of the function $Q(\zeta)$ allow us to make
the following conclusions. At each regular interval the equation
\be Q(\zeta) = \alpha \label{spec} \ee
has one and only one solution. We denote it by $E_k(\alpha)$ where
the index refers to the right endpoint of the interval in
question. If $|b|$ is half-integer the equation (\ref{spec}) has
at the special interval a solution for any $\alpha\in\R$,
otherwise a solution exists there if and only if the inequality
$$ 4\pi\alpha < \psi({1\over 2} +\!b) + \psi({1\over 2}-\!b) +
2\gamma - 2\ln 2a $$
is valid; if the solution at the special interval exists, it is
unique and we shall denote it by $E_{n_0+1}(\alpha)$.

It follows from (\ref{krein}) that the discrete spectrum of
$H_{w,\alpha}$ consists exactly of all solutions of the equation
(\ref{spec}) in the interval $(-\infty,b^2/a^2)$. Any such
solution $E_k(\alpha)$ is at that a simple eigenvalue; the
corresponding normalized eigenfunction $\Psi_k(z;w,\alpha)$ is
given by
$$ \Psi_k(z;w,\alpha) = c_k\, G^0(z;w,E_k(\alpha)) $$
with the normalization factor
$$ c_k = \left\lbrack \left.{\pd Q\over \pd\zeta}
\right|_{\zeta=E_k(\alpha)} \right\rbrack^{-1/2}. $$
For our future purpose it is important that $E_k(\alpha)$ is
independent of $w$, and therefore it remains to be a simple
isolated eigenvalue as the position of the point perturbation is
changed.


\setcounter{equation}{0}
\section{The Berry phase}

We shall now realize our aim of finding the Berry phase for an
adiabatic evolution of our system in the parameter space
$\HHH^2_a\ni w$. Let us compute the corresponding Berry potential.
Since $\alpha$ and the level index $k$ are kept fixed in the
following we drop them from the notations. We first remark that
the eigenfunction $\Psi$ can be written in the form
\be \Psi(z;w) = \left(- {z\!-\!\bar z' \over \bar z \!-\!z'}
\right)^b \phi(\sigma(z,w))\,, \label{ef} \ee
where the function $\phi$ is real-valued. The derivatives of the
first factor with respect to $u=\Re w$ and $v=\Im w$ are
\ba {\pd\over \pd u} \left(- {z\!-\!\bar z' \over \bar z \!-\!z'}
\right)^b &\!=\!& -b \left(- {z\!-\!\bar z' \over \bar z \!-\!z'}
\right)^{b-1} {z\!-\!\bar z\!+\!w\!-\!\bar w\ \over (\bar
z\!-\!w)^2}\,, \label{uder} \\ {\pd\over \pd v} \left(-
{z\!-\!\bar z' \over \bar z \!-\!z'} \right)^b &\!=\!& -bi \left(-
{z\!-\!\bar z' \over \bar z \!-\!z'} \right)^{b-1} {z\!+\!\bar
z\!-\!(w\!+\!\bar w)\ \over (\bar z\!-\!w)^2}\,.\ea
The last relations in combination with (\ref{ef}) gives
\ba  \langle \Psi|{\pd\over \pd v} \Psi \rangle &\!=\!& -2bi
\int_{\HHH^2_a} {x\!-\!u \over (x\!-\!u)^2\! +(y\!+\!v)^2} \,
[\phi(\sigma(z,w))]^2 d\mu_a(z) \nonumber \\ && + \int_{\HHH^2_a}
\phi(\sigma(z,w)) \,{\pd\over \pd v}\, \phi(\sigma(z,w))
d\mu_a(z)\,. \label{v-comp} \ea
Since $\phi$ is real-valued and $\int_{\HHH^2_a}
[\phi(\sigma(z,w))]^2 d\mu_a(z) = 1$ holds for all $w\in
\HHH^2_a$, the second integral in (\ref{v-comp}) vanishes. Using
the substitution $z\!-\!u \to z$ in the first one, we get the
expression
\be  \langle \Psi|{\pd\over \pd v} \Psi \rangle = -2bi
\int_{\HHH^2_a} {x \over x^2\! +(y\!+\!v)^2} \,
[\phi(\sigma(z,iv))]^2 d\mu_a(z) \,. \label{v-comp2} \ee
By (\ref{sig}) the function $\sigma(x,iv)$ is even with respect to
$x$, hence the integrated function in (\ref{v-comp2}) is odd and
$$  \langle \Psi|{\pd\over \pd v} \Psi \rangle = 0 \,. $$

Let us turn to the $u$-component of the Berry potential. Since
$\phi$ is real-valued, we infer from (\ref{ef}) and (\ref{uder})
\begin{eqnarray*}  \langle \Psi|{\pd\over \pd u} \Psi \rangle &\!=\!& -2bi
\int_{\HHH^2_a} {y\!+\!v \over (x\!-\!u)^2\! +(y\!+\!v)^2} \,
[\phi(\sigma(z,w))]^2 d\mu_a(z) \nonumber \\ &\!=\!& -2bi
\int_{\HHH^2_a} {y\!+\!v \over x^2\! +(y\!+\!v)^2} \,
[\phi(\sigma(z,iv))]^2 d\mu_a(z) \,. \end{eqnarray*}
Another substitution, $z\to vz$, yields
$$  \langle \Psi|{\pd\over \pd u} \Psi \rangle = -\, {2bi\over v}
\int_{\HHH^2_a} {y\!+\!1 \over x^2\! +(y\!+\!1)^2} \,
[\phi(\sigma(z,i))]^2 d\mu_a(z) \,, $$
and since $\sigma(z,i) = [x^2\!+(y\!+\!1)^2]/4y$, we have
$$  \langle \Psi|{\pd\over \pd u} \Psi \rangle = -\, {bi\over 2v}
\int_{\HHH^2_a} {y\!+\!1 \over y\sigma(z,i)} \,
[\phi(\sigma(z,i))]^2 d\mu_a(z) \,. $$
We have remarked already that $\sigma$ is independent of $a$; then
it follows from (\ref{mu}) that
\be  \langle \Psi|{\pd\over \pd u} \Psi \rangle = -\, {bi a^2\over
2v} \int_{\HHH^2_a} {y\!+\!1 \over y\sigma(z,i)} \,
[\phi(\sigma(z,i))]^2 d\mu_1(z) \,. \label{u-comp} \ee
To evaluate the last integral we pass to the polar coordinate
system centered at the point $i$ putting $r=d_1(z,i)$. Then
$$ y^{-1} = \cosh r + \sinh r\, \cos 2\varphi\,, \qquad d\mu_1(z)
= \sinh r\, dr\,d\varphi \,, $$
where $\varphi$ is the polar angle \cite{Ter}. Since $\sigma(z,i)
= \cosh^2{r\over 2}$ we can rewrite the r.h.s. of (\ref{u-comp})
as
$$ -\, {bi a^2\over 2v} \int_0^{\infty}\! dr \int_0^{2\pi}
\!d\varphi\, (1+\cosh r + \sinh r\, \cos 2\varphi)\, {\sinh r\over
\cosh^2{r\over 2}}\, \left\lbrack\phi\left(\cosh^2{r\over
2}\right)\right\rbrack^2\,. $$
Using $1+\cosh r= 2\,\cosh^2{r\over 2}$ and integrating over
$\varphi$ we find
$$ \langle \Psi|{\pd\over \pd u} \Psi \rangle = -\, {2\pi bi
a^2\over v} \int_0^{\infty} \left\lbrack\phi\left(\cosh^2{r\over
2}\right)\right\rbrack^2 \sinh r\, dr\,. $$
On the other hand,
\begin{eqnarray*}  \lefteqn{ 2\pi a^2 \int_0^{\infty}
\left\lbrack\phi\left(\cosh^2{r\over 2}\right)\right\rbrack^2
\sinh r\, dr = a^2 \int_0^{\infty}\! dr \int_0^{2\pi} \!d\varphi
\left\lbrack\phi\left(\sigma(z,i)\right)\right\rbrack^2 \sinh r }
\nonumber \\ && = a^2 \int_{\HHH^2_1} [\phi(\sigma(z,i))]^2
d\mu_1(z) = \int_{\HHH^2_a} [\phi(\sigma(z,i))]^2 d\mu_a(z) =
\|\Psi(\cdot\,;i)\|^2 =1 \,, \end{eqnarray*}
so finally we arrive at the expression
$$ \langle \Psi|{\pd\over \pd u} \Psi \rangle = -\, {ib \over
v}\,. $$
Hence the Berry potential,
$$ \V(w) =i\, \langle \Psi(\cdot\,;w)|\nabla_w \Psi(\cdot\,;w)
\rangle $$
of our system is of the form
$$ \V(w) = \left( {b\over v}, 0 \right) = \left( {Ba^2\over v}\,,
0 \right) $$
i.e. similarly to the case of the Euclidean plane it coincides
with the vector potential $\A$ if we express $\V$ as a 1-form
${Ba^2\over v}\,du$.

Let now $\CC$ be a smooth closed contour in the Lobachevsky plane
$\HHH_a^2$, then the Stokes formula yields the sought expression
for the Berry phase $\gamma(\CC)$:
\be \gamma(\CC) = \int_{\CC} {Ba^2\over v}\,du = \int\!\!\int_S
{Ba^2\over v}\,du\wedge dv = BS = {2\pi\Phi_{\CC}\over \Phi_e} \,,
\label{berry} \ee
where $\Phi_{\CC}$ is the total flux of the field $\B$ through the
area $S$ encircled by the loop $\CC$. The relation (\ref{berry})
is the main result of this letter.

Notice that in distinction of the spectrum of the Hamiltonian
$H_{w,\alpha}$ the Berry phase depends neither on the curvature
$R$ nor on the coupling parameter $\alpha$. In particular,
$\gamma(\CC)$ is independent of the energy of the considered
particle in the zero-range potential well which confines it.


\subsection*{Acknowledgment}

This research has been partially supported by GAAS and Czech
Ministry of Education under the contracts 1048801 and ME099. The
last named author is also very grateful to the DFG (Grant 436 RUS
113/572/1) and RFFI (Grant No 98-01-03308) for a financial
support.


\end{document}